\documentclass[twocolumn]{aastex63}
\usepackage{multirow}

\usepackage{amsmath}
\usepackage{calligra}
\usepackage{xspace}

\usepackage{color}

\definecolor{chmagenta}{rgb}{0.54, 0.17, 0.88}

\newcommand{\posydon}{\texttt{POSYDON}\xspace}
\newcommand{\mesa}{\texttt{MESA}\xspace{}}

\newcommand{\SukhboldWtwenty}{Sukhbold16-engineW20}
\newcommand{\SukhboldNtwenty}{Sukhbold16-engineN20}
\newcommand{\PattonSukhboldWtwenty}{PattonSukhbold20-engineW20}

\DeclareMathAlphabet{\mathcalligra}{T1}{calligra}{m}{n}
\DeclareFontShape{T1}{calligra}{m}{n}{<->s*[2.2]callig15}{}

\newcommand{\alphaCE}{$\alpha_{\mathrm{CE}}$}

\newcommand{\be}{\begin{enumerate}}
\newcommand{\ee}{\end{enumerate}}

\shorttitle{}
\shortauthors{Gallegos-Garcia et al.}

\begin{document}

\title{The Mass-Ratio Distribution of the Low-Mass Binary Black Hole Subpopulation: \\ A Natural Outcome of Isolated Binary Evolution} 

\author[0000-0003-0648-2402]{Monica Gallegos-Garcia}
\affiliation{Harvard Society of Fellows, 78 Mount Auburn Street, Cambridge, MA 02138, USA}
\affiliation{Center for Astrophysics \textbar{} Harvard \& Smithsonian, 60 Garden St. Cambridge, MA, 02138, USA}

\author[0000-0002-7322-4748]{Anarya Ray}
\affiliation{Center for Interdisciplinary Exploration and Research in Astrophysics (CIERA),1800 Sherman, Evanston, IL 60201, USA}
\affiliation{NSF-Simons AI Institute for the Sky (SkAI),172 E. Chestnut St., Chicago, IL 60611, USA}

\author[0000-0001-9236-5469]{Vicky Kalogera}
\affiliation{Center for Interdisciplinary Exploration and Research in Astrophysics (CIERA),1800 Sherman, Evanston, IL 60201, USA}
\affiliation{NSF-Simons AI Institute for the Sky (SkAI),172 E. Chestnut St., Chicago, IL 60611, USA}
\affiliation{Department of Physics and Astronomy, Northwestern University, 2145 Sheridan Road, Evanston, IL 60208, USA}

\author[0000-0002-6842-3021]{Max Briel}
\affiliation{Département d’Astronomie, Université de Genève, Chemin Pegasi 51, CH-1290 Versoix, Switzerland}
\affiliation{Gravitational Wave Science Center (GWSC), Université de Genève, CH1211 Geneva, Switzerland}

\author[0000-0002-0147-0835]{Michael Zevin}
\affiliation{The Adler Planetarium, 1300 South DuSable Lake Shore Drive, Chicago, 60605, IL, USA}
\affiliation{Center for Interdisciplinary Exploration and Research in Astrophysics (CIERA),1800 Sherman, Evanston, IL 60201, USA}
\affiliation{NSF-Simons AI Institute for the Sky (SkAI),172 E. Chestnut St., Chicago, IL 60611, USA}

\author[0000-0002-6064-388X]{Abhishek Chattaraj}
\affiliation{Department of Physics, University of Florida, 2001 Museum Rd, Gainesville, FL 32611, USA}

\author[0000-0002-0031-3029]{Zepei Xing}
\affiliation{Center for Interdisciplinary Exploration and Research in Astrophysics (CIERA),1800 Sherman, Evanston, IL 60201, USA}

\author[0000-0001-5261-3923]{Jeff J. Andrews}
\affiliation{Department of Physics, University of Florida, 2001 Museum Rd, Gainesville, FL 32611, USA}
\affiliation{Institute for Fundamental Theory, 2001 Museum Rd, Gainesville, FL 32611, USA}

\author[0000-0001-6692-6410]{Seth Gossage}
\affiliation{Center for Interdisciplinary Exploration and Research in Astrophysics (CIERA),1800 Sherman, Evanston, IL 60201, USA}
\affiliation{NSF-Simons AI Institute for the Sky (SkAI),172 E. Chestnut St., Chicago, IL 60611, USA}

\author[0000-0003-1749-6295]{Philipp M. Srivastava}
\affiliation{Center for Interdisciplinary Exploration and Research in Astrophysics (CIERA),1800 Sherman, Evanston, IL 60201, USA}
\affiliation{NSF-Simons AI Institute for the Sky (SkAI),172 E. Chestnut St., Chicago, IL 60611, USA}
\affiliation{Electrical and Computer Engineering, Northwestern University, 2145 Sheridan Road, Evanston, IL 60208, USA}

\begin{abstract}

The observed merging binary black hole population is increasingly consistent with being composed of a mixture of subpopulations, each likely the result of different formation mechanisms. 
In particular, the low-mass subpopulation, with primary black hole masses below $\approxeq 15\,M_{\odot}$, has been attributed to mergers formed through isolated binary evolution. 
We use the binary population synthesis code \posydon{} to study the mass ratio $(q)$ distribution of binary black hole mergers from isolated binary evolution and compare to the observed low-mass subpopulation. 
We explore variations in supernova remnant prescriptions, common-envelope efficiency, and black hole accretion efficiency. 
We find that our models have a preference for asymmetric $q$, most with a broad peak near $q\approxeq0.5-0.7$, and a near-equal-mass component whose relative strength varies across models.
This resultant $q$ distribution is consistent with the $q$ distribution of the low-mass subpopulation observed with gravitational waves. 
We find that for the majority of models these features arise from physically distinct formation subchannels: the asymmetric peak reflects contributions from common-envelope and stable mass-transfer systems, while the near-equal-mass component traces double-core common envelope and contact systems.
We conclude that the features in the $q$ distribution of the observed low-mass subpopulation emerge naturally from isolated binary evolution across a range of model assumptions.
As the gravitational-wave catalog continues to grow, the relative strength of these $q$ features will provide an increasingly powerful diagnostic of isolated binary evolution and its formation subchannels.
\end{abstract}


\section{Introduction}
The growing number of gravitational-wave (GW) detections has opened new pathways to uncover the origins of merging binary black holes (BBH).
Where early on, the observed BBH merger population was studied as a {\it single} population, the increase of detections has allowed statistical methods to probe substructure and identify {\it subpopulations} within the full catalog of BBH mergers \citep{Li2022, Godfrey2023, Sadiq:2023zee, Afroz:2024fzp, Antonini:2024het, Galaudage:2024meo, Banagiri2025, Ray_BGP_2025, Sridhar2025, Wang:2025nhf, Afroz:2025ikg, Li:2025iux, Antonini:2025ilj, Tong:2025wpz, LVK_O4_2025, Roy2025, Farah:2026jlc, Vijaykumar:2026zjy, Galaudage:2026opk, Ray_subpop_2026, Flanagan:2026ayy, Flanagan:2026btu, Cheng:2026bpc, Alvarez-Lopez:2026ymo, Rinaldi:2026nyb, Li:2026iae, Zeeshan:2026mpr, Guttman:2026cnv, Padhyegurjar:2026scg, Padhyegurjar:2026slt, Ray:2026qer}. 
This multi-component framework is particularly powerful for testing theoretical predictions: since the diversity of BBH properties suggests that multiple formation pathways are at play \citep[e.g.][]{Zevin2021,Colloms2025}, the identification of subpopulations with distinct features may be a way to reveal where, across the full catalog, certain formation channels dominate the formation of BBH mergers.
 
By contrast, the longstanding paradigm in BBH formation simulations has primarily been to consider the formation of BBH mergers for one formation channel and compare it to the full observed population.
The distinct formation pathways considered can be broadly separated into the following \citep[for recent reviews spanning multiple channels, see][]{Mapelli2021,MandelBroekgaarden2022,MandelFarmer2022}: i) isolated binary evolution, in which two massive stars evolve through episodes of mass transfer (MT), a common-envelope (CE) phase, and/or chemically homogeneous evolution (CHE) to produce a tight compact object (CO) binary that will merge within a Hubble time, ii) dynamically formed binaries, in which BHs pair through interactions either in dense stellar environments such as globular clusters and nuclear star clusters, gaseous environments in AGN disks, or triple systems \citep[see also][]{GerosaFishbach2021,ArcaSedda2023,Kremer2026}, and iii) binary mergers of primordial origin~\citep{Nakamura:1997sm, Ioka:1998nz, Bird:2016dcv, Clesse:2016vqa, Sasaki:2016jop, Mandic:2016lcn, Raidal:2017mfl, Ali-Haimoud:2017rtz, Raidal:2018bbj, Vaskonen:2019jpv, Jedamzik:2020ypm, Huang2024GW230529, Aljaf2025, ElBouhaddouti2026, Yuan2025GW231123, DeLuca2026GW231123, Aljaf2026MergerRate, AndresCarcasona2026}.

A more targeted comparison is now possible: instead of asking whether one formation channel can explain the entire catalog, we can identify which subpopulations are most compatible with theoretical predictions from a given formation mechanism~\citep[e.g.][]{Ray_subpop_2026, Galaudage:2026opk}.
The latest analysis of the GW catalog has shown that predicted signatures of different mechanisms, such as in mass-ratios~($q$) and effective inspiral spins~($\chi_{\mathrm{eff}}$), may, in fact, be emerging more strongly in certain subpopulations than others \citep{LIGOScientific:2026ctl, Ray_subpop_2026, Cheng:2026bpc, Galaudage:2026opk, Alvarez-Lopez:2026ymo, Flanagan:2026ayy, Flanagan:2026btu, Guttman:2026cnv, Ray:2026qer}.% \textbf{(add more citations)}

In this study, our focus is on the $q$ distribution. Previous analyses have shown that instead of a single $q$ distribution across all masses, the data prefer that particular mass ranges have distinct $q$ distributions~\citep{Sadiq:2021fin, Godfrey2023, Sadiq:2023zee, Ray_BGP_2025, Banagiri2025, Sridhar2025, Ray_subpop_2026, Galaudage:2026opk}. While significant model-dependence exists on the exact shape of these mass-dependent $q$ distributions~\citep{LIGOScientific:2026ctl},  most analyses are in agreement that the $q$ distribution at low-mass~($m_1 \approxeq 8-15M_{\odot}$) has support over a broad range ($q\approxeq0.6-1.0$) as opposed to a clear preference for equal-mass~($q\approxeq 1$) in the mid-mass~($m_1 \approxeq 25-45M_{\odot}$) subpopulation, or unequal mass in the intermediate~\citep[$m_1 \approxeq 18-23M_{\odot})$][]{Flanagan:2026ayy, Alvarez-Lopez:2026ymo, Guttman:2026cnv, Ray:2026qer} and high-mass $(m_1 \gtrsim 60M_{\odot})$  subpopulations~\citep{Banagiri2025, Sridhar2025, Ray_subpop_2026, Guttman:2026cnv, Ray:2026qer}. 

On the theoretical side, a critical question is whether the $q$ distributions predicted for the various formation channels can indeed be associated with any of the inferred subpopulations.
Predictions for dynamical assembly of first-generation merging BBHs in dense clusters~\citep{Mapelli:2021gyv, Chattopadhyay:2023pil, Rodriguez:2021qhl} strongly favor equal-mass systems due to mass segregation which causes heavier objects to sink to the center of the cluster~\citep[e.g.][]{Rodriguez:2016vmx, Antonini2023}, and strong gravitational interactions between three or more bodies that preferentially lead to the more massive components forming a bound binary and ejecting less massive components~\citep{Heggie:1997gq}. When compared to the observed subpopulations, this formation channel, while consistent with the mid-mass range, is unlikely to dominate the low-mass subpopulation, which has substantial support for a broader range, ($q\approxeq0.6-1.0$). Contributions from other dynamical pathways such as triple systems~\citep{Antonini:2017ash, Rodriguez:2018jqu, Liu:2018nrf, Stegmann:2025zkb}, hierarchical mergers, and binaries in AGN disks can also have higher support for unequal-mass systems ~\citep{Martinez:2022TripleMassRatio,Su:2021TripleMassRatio,Trani:2022TripleMassRatio,Dorozsmai:2025Triples, Fitchett:1983qzq,PortegiesZwart:1999nm,Favata:2004wz,Gonzalez:2006md,Lousto:2009ka,Gerosa:2018qay,Mahapatra:2021hme, Zevin:2022bfa, Tagawa:2020dxe, McKernan:2021nwk, Li:2022cul,Mckernan:2017ssq, Santini:2023ukl, McKernan:2023xio, McKernan:2024kpr, Cook:2024ajp, Fabj:2025vza}, although these predictions are susceptible to uncertainties in treatments of binary evolution and disk physics. 

For isolated binary evolution, most predictions come from rapid population synthesis codes built on analytic fits to single-star models. 
These predict a preference for near-equal-mass BBH mergers \citep[e.g.][]{Belczynski2016, Giacobbo2018, Bouffanais2021}, broad distributions extending to $q \approxeq 0.2$, and asymmetric features in $q$, the latter often when the SMT channel and mass-ratio reversal are emphasized; the location and shape of these features vary widely across studies \citep{Dominik2012, Spera2019, Neijssel2019, Bavera2020, Olejak2021,Shao2021, ZevinBavera2022, Broekgaarden2022, vanSon2022_redshift, Olejak2024, Torniamenti2024, Chen2026, Smith2026}. 
Other predictions apply to individual formation subchannels. 
For example, \citet{vanSon2022_redshift} identify an asymmetric peak near $q \approxeq 0.7$ for the SMT channel alone, and show that the width of the peak for this channel, when considering both BBH and neutron star–black hole mergers, depends on the assumed MT physics \citep{vanSon_2022_peaks}.
For CHE, $q$ is predicted to cluster tightly near unity \citep{deMink2016, Marchant2016}. 
Studies that use detailed binary models for part of the evolution show a comparable spread, favoring a clustering around $q \approxeq 0.7$ across the full population \citep{Bavera2021} or a broad distribution down to $q \approxeq 0.2$ \citep[e.g.][]{Ghodla2022, Briel2023}. 
Finally, results from detailed binary models, those that solve the equations for stellar structure during binary evolution, are so far restricted to a single subchannel, such as SMT, or adopt progenitor conditions that do not span a complete astrophysical population \citep{Briel2026, Klencki2026, Xu2025}.  
No prediction to date therefore combines detailed binary models with an astrophysical population sampling all formation subchannels.

In this Letter, we explore the $q$ distribution of isolated BBH mergers with \posydon{}. 
\posydon{} uses binary evolution \mesa{} simulations that self-consistently evolve both stars through all binary interactions, making it among the most physically detailed binary population synthesis frameworks currently available for large-scale population studies.
Rather than performing an exhaustive calibration to observations, we identify the broad features of isolated binary evolution that persist across reasonable variations in uncertain physics. 
We show that isolated binary evolution displays a preference for asymmetric $q$ with an equal-mass contribution, consistent with the observed low-mass component of BBH mergers. 

This Letter is structured as follows. 
In Section~\ref{sec:methods} we describe the \posydon{} simulations. 
In Section~\ref{sec:results} we present the $q$ distributions across our models and compare our predictions to observations. 
We discuss caveats in Section~\ref{sec:caveats} and summarize our conclusions in Section~\ref{sec:conclusions}.
\\
\\

\section{Method}\label{sec:methods}

We use the binary population synthesis code \posydon{} version 2.1.0  \citep{Fragos2023, Andrews2025} to simulate the evolution of binary systems. 
Here we highlight the physics most relevant in the formation of BBH mergers.
For an additional detailed description of the assumptions to stellar and binary physics focused on BBHs merging via SMT see \cite{Briel2026}.

Our \posydon{} binary evolution simulations are constructed from grids of single- and binary-star models, computed using the 1D stellar evolution code \mesa{} \citep{Paxton2011, Paxton2013, Paxton2015, Paxton2018, Paxton2019, Jermyn2023}.
Our simulations span eight metallicities: $10^{-4}$, $ 10^{-3}$, $10^{-2}$, $0.1$, $0.2$, $0.45$, $1.0$, and $2.0 Z_{\odot}$, with $Z_{\odot}=0.0142$, and BH progenitor stars are evolved up to core carbon depletion. 
Using these grids of \mesa{} simulations, we sample $10^6$ binaries per metallicity from a \citet{Kroupa2001} initial mass function and a flat $q$ distribution, with a binary separation sampled from a log uniform distribution between $5-10^5\ R_{\odot}$. 
The binary systems are further weighted according to the star formation and metallicity evolution of the IllustrisTNG-100 simulation \citep{Springel+18, Nelson+18, Pillepich+18, Naiman+18, Marinacci+18}. 
While our focus is to compare to the observed low-mass component, we consider all BBH mergers in our simulations since we find that low-mass BBHs with $m_1\lesssim 25 M_{\odot}$ dominate our simulated populations.

\subsection{Stellar Physics}
The stellar physics assumptions regarding winds and SNe physics are the following.
For hot, hydrogen-rich stars, \posydon{} uses the stellar wind mass loss from \citet{Vink2000} with a $(Z/Z_\odot)^{0.68}$ dependence on metallicity \citep{Vink2001}. For Wolf-Rayet-like winds from hot helium-rich stars, the models of \citet{Nugis2000} are used. 
For stars reaching the Humphreys-Davidson limit, an enhanced mass loss of $10^{-4}\ M_{\odot} {\mathrm{yr}}^{-1}$ is applied \citep{Belczynski10a}.
For the formation of BHs, we vary between three SNe prescriptions: i)~\cite{Sukhbold2016} using engine N20 (\SukhboldNtwenty{}), ii)~\cite{Sukhbold2016} using engine W20 (\SukhboldWtwenty{}), and iii)~\cite{Patton2020} using engine W20 (\PattonSukhboldWtwenty{}). 
All three determine explodability from the pre-collapse core structure and none imposes a lower mass gap by construction.
In practice, BHs below $\sim 5\ M_{\odot}$ are rare in our populations.
For SNe natal kicks, we apply a mass-scaled kick, rescaling kick magnitudes drawn from a Maxwellian distribution with dispersion $\sigma=265 {\mathrm{\ km \ s}}^{-1}$. 

\subsection{Binary Evolution Physics}
%binary physics 
The binary physics assumptions regarding MT treatments, angular momentum transport onto a star, and MT stability are as follows.
For MT between two stars, \posydon{} uses the \texttt{contact} scheme in \mesa{} for Roche lobe overflow on the main-sequence, allowing for both stars to fill their Roche lobes simultaneously and form a contact system.
For post main-sequence stars, \posydon{} uses the \texttt{kolb} MT scheme \citep{Kolb1990}.
For the treatment of angular momentum transfer onto a stellar companion during MT, \posydon{} follows \cite{deMink2013} for the specific angular momentum carried by the accreted material, which can spin up the accretor star. 
If this MT causes the accretor to reach near critical rotation, \posydon{} implements rotational-limited accretion, where the increase in stellar rotation enhances the mass loss through stellar winds causing the star to lose angular momentum and remain below critical rotation. 

Mass-transfer stability in \posydon{} is determined self-consistently for each binary model. 
For binaries with BH accretors, MT is considered unstable and leads to a CE if 1) the MT exceeds either $0.1\ M_{\odot} {\mathrm{yr}}^{-1}$ or 2) the system experiences outflow through the outer Lagrangian point ($L_2$ if the accretor is less massive than the donor, and $L_3$ otherwise). 
MT rates of $0.1\ M_{\odot} {\mathrm{yr}}^{-1}$ are expected to lead to a runaway process, causing mass loss rates to exceed both thermal and dynamical timescale MT. 
We therefore assume these binary systems become dynamically unstable.
For outflow from $L_2$, it is expected that the ejected material will remove significant angular momentum given its location away from the center of mass of the system. 
This significant loss of angular momentum is believed to then cause a rapid inspiral leading to instability.
If one of these conditions is reached, the \mesa{} model is terminated and \posydon{} calculates the evolution of CE.

There are two key differences in the treatment of CE in \posydon{} compared to the standard energy budget formalism commonly used, also known as the $\alpha-\lambda$ prescription \citep[for a review see][]{Ivanova2020,TaurisvandenHeuvel2023}: First, the $\alpha-\lambda$ prescription traditionally involves adopting assumptions for several free parameters: $\lambda$ which adjusts the binding energy of the envelope, \alphaCE{} which treats the efficiency of reducing the orbital separation through this process, and an assumption for the location for the core-envelope boundary.
With \posydon{} simulations, however, we can eliminate one of these free parameters. 
We use the density profiles calculated with \mesa{} at the onset of CE to self-consistently determine the binding energy of the envelope to the star.
In our simulations, we must still adopt a CE efficiency \alphaCE{} and a core-envelope boundary.
Second, we use the two-step process to calculate the orbital evolution \citep[see also][]{Hirai2022}, motivated by detailed studies of CE \citep{Ivanova2011,Fragos2019,Marchant2021}. The first step applies the energy formalism described above, with a core–envelope boundary defined by the mass shell where the central hydrogen mass fraction drops below 30\%. 
Assuming the outcome is a detached system, the second calculation treats the remaining envelope layer starting from the core-envelope boundary to the mass shell where the fraction of hydrogen mass drops below 1\%; we assume that this layer is removed via non-conservative MT with the specific angular momentum of the accretor.

Contrary to the default assumptions in \posydon{}, for this study we decide to exclude CE systems triggered by the $L_2$ overflow condition \citep[see Section 4.2.4. in][]{Fragos2023}.
Previous studies have shown that this MT can remain stable despite a greater loss of angular momentum \citep[e.g.][]{Marchant2021}.
We discuss this further in Appendix~\ref{sec:L2_CE_appendix} and show that if included, these systems exhibit similar properties to the CE systems triggered by the $\dot{M}$ criteria and therefore do not significantly change the qualitative results of our study. 

% mass transfer 
For MT physics onto a BH, we explore the standard assumption of Eddington-limited accretion. 
In addition to the publicly-available grids,\footnote{\url{https://zenodo.org/records/15194708}} we use an alternative BH accretion efficiency that was implemented into the grids of \posydon{}.
This prescription is informed by general relativistic radiation magnetohydrodynamic (GRRMHD) simulations \citep{Kwan2026}, and can allow for efficiencies of $10-30\%$ (see \cite{Xing2025} for a description of the MT prescription in \posydon{}).

In addition to the three SNe prescription variations, we explore variations to \alphaCE{} including \alphaCE{}$=0.5, 1.0, 2.0$, and vary MT between Eddington-limited and the GRRMHD-informed prescription. 
Table~\ref{table:models} in the Appendix lists the 15 models in this study.

\section{Results} \label{sec:results}
\subsection{Mass Ratio Distribution}\label{sec:primary_mass_q_distributions}

\begin{figure*}
    \centering
\includegraphics[width=0.95
\textwidth]{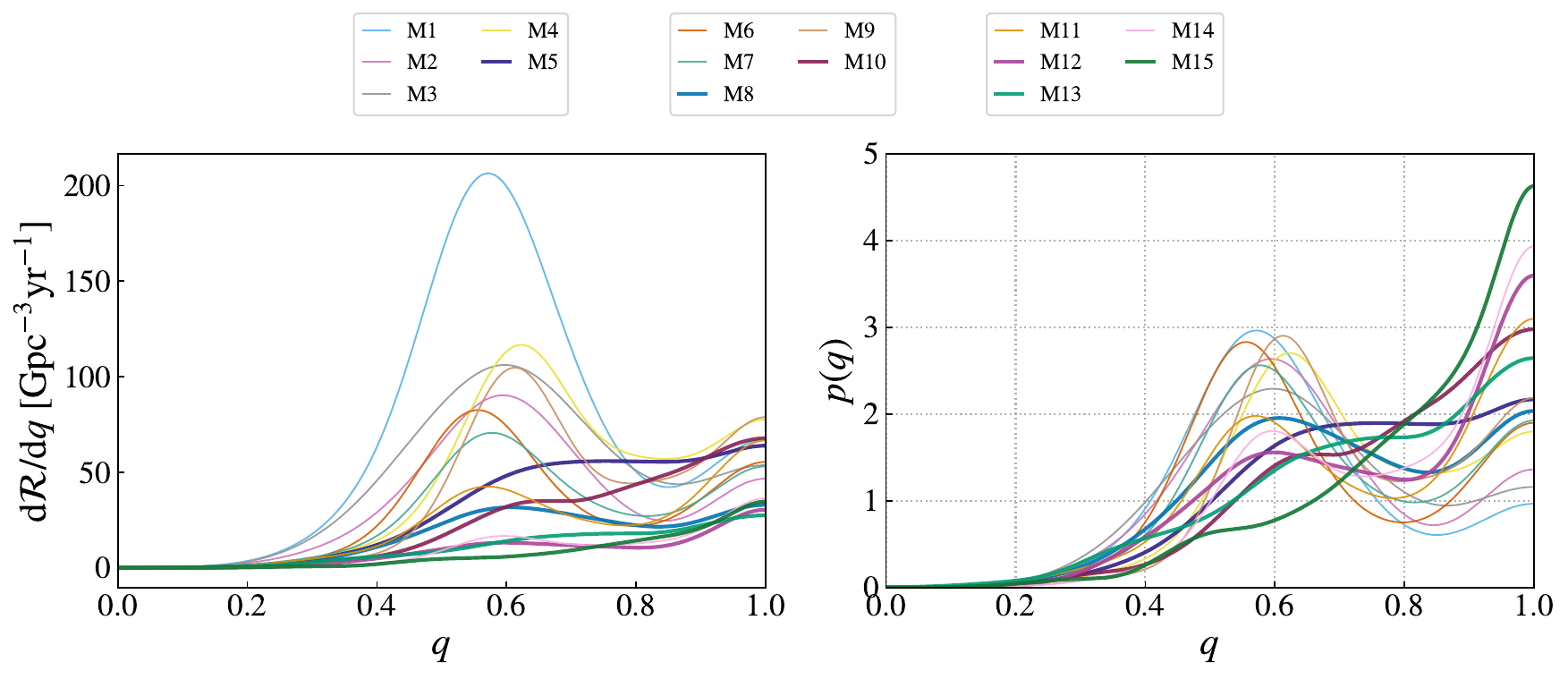}
    \caption{Mass-ratio distribution of merging BBHs within the local Universe, $z<0.5$, for all models. 
    \textit{Left:} merger rate density, \textit{Right:} probability density. The latter highlights the robustness of the shape across models independent of rate.
    We find that for a range of assumptions our models display similar features: a contribution at both equal and unequal $q$, with a range of smoothness across the peaks.}
    \label{fig:q_all_models}
\end{figure*}

In this section, we present the $q$ distributions for BBHs merging within $z~<~0.5$.
In Figure~\ref{fig:q_all_models}, we show the $q$ distribution as both the merger rate density $\mathrm{d}\mathcal{R}/\mathrm{d}q$ (left panel) and the normalized probability density $p(q)$ (right panel).
The merger rate density illustrates the range of predicted rates across models, while the normalized probability density shows the shape of the $q$ distribution across models.
Despite the range in rates, we find similar features in the shape of the $q$ distributions: the majority of our models display either a narrow or broad peak near asymmetric values of $q\approxeq0.5$--$0.7$ and an equal-mass component whose contribution varies with model assumptions.
The shape between these two features varies from a drop, hence a double-peaked distribution, to a plateau, hence a flatter $q$ distribution.

The shape of the $q$ distribution is determined by the peak locations of formation subchannels.
For the majority of our models, we find that the asymmetric peak in the full distribution reflects the combined contribution of systems formed through standard CE evolution and systems without an unstable CE phase formed through only SMT. 
Both formation subchannels tend to peak in this region of the $q$ distribution, although there is more variation in location of the peak for systems formed through SMT. 
The secondary near-equal-mass contribution arises primarily from double-core CE systems (in which two post-main-sequence stars simultaneously overflow their Roche lobes, triggering a CE event where both stellar cores inspiral within a shared envelope) and contact SMT systems (in which two hydrogen-rich stars overflow their Roche lobes simultaneously during a SMT episode, entering a contact phase). 
This behavior may be explained by i) initial near-equal masses are necessary for evolution through double-core CE and contact SMT, and we find that this $q\approxeq1$ is preserved throughout the evolution, ii) SMT and CE systems tend to have asymmetric mass loss, where one component is stripped more than the other, which may result in an asymmetric $q$, though there is more variation for systems formed through SMT. 
An example of the $q$ distributions of these formation subchannels can be found in Figure~\ref{fig:q_with_L2CE_appendix} in the Appendix where we split the full population into all subchannels, including CE systems triggered through the $L_2$ criteria.
A detailed study of the locations of these formation subchannels and how their relative contributions and locations depend on stellar and binary physics assumptions will be presented in Gallegos-Garcia et al. 2026, in prep., where we explore these mechanisms directly. 

We find that the local BBH merger rate varies across models from {$\approxeq~54-1860\,\mathrm{Gpc}^{-3}\,\mathrm{yr}^{-1}$ at $z=0.2$.
Our lowest-rate model, M13, lies just above the upper bound of the 95\% credible interval inferred from the most recent catalog by the LIGO-Virgo-KAGRA (LVK) collaboration, $\approxeq~27-50\,\mathrm{Gpc}^{-3}\,\mathrm{yr}^{-1}$ at $z=0.2$ \citep{LIGOScientific:2026wfs}, while our highest-rate model exceeds the 95\% credible interval by a factor of $\approxeq 37$.
Reproducing the observed merger rate remains an active area of research, and binary population synthesis models generally struggle to match it \citep{MandelBroekgaarden2022}, even when adopting the most up-to-date cosmic star formation rate models \citep[e.g.][]{Boco2026,Sgalletta2025}, though see \citet{Broekgaarden2026}.
We conclude that the general qualitative features of the $q$ distribution are robust across our model variations, even as the predicted merger rate varies.
 
\subsection{Comparison to Observations}\label{sec:comparison_to_obs}

\begin{figure}
\includegraphics[width=0.49
\textwidth]{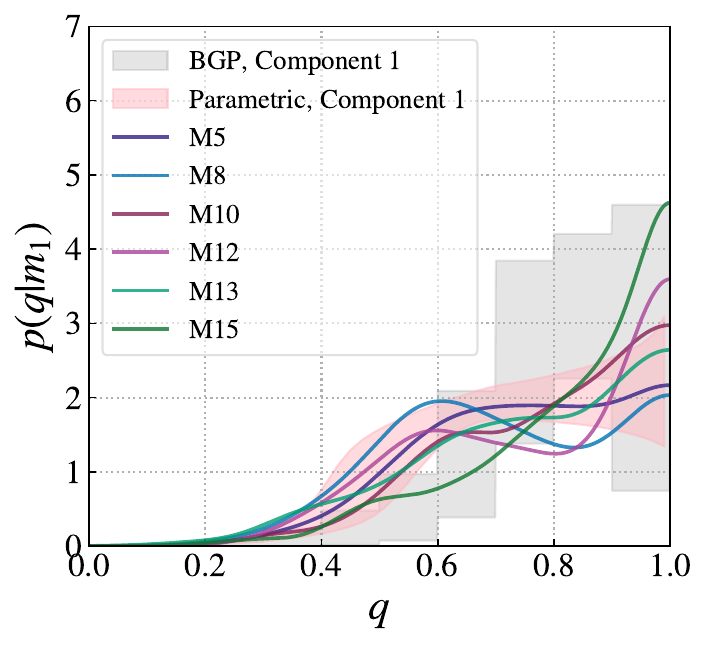}
    \caption{Comparison of our isolated binary evolution models to the 90\% credible interval inferred from GWTC-5.0, using the data-driven binned Gaussian process approach in \citet{Sridhar2025} (gray shaded region) and the parameterized mixture models of \citet{Ray_subpop_2026} (pink shaded region).  
    The colored lines show a selected sample of our model predictions. }
    \label{fig:comparison_to_obs}
\end{figure}

Analysis of the latest GW catalogs has revealed a complex underlying population of BBH mergers. 
Several studies have characterized subpopulations of systems and searched for direct imprints of binary evolution.
In this section, we compare our theoretical results for $q$ distributions of BBH mergers to two data-driven inferences of the GWTC-5.0 \citep{LIGOScientific:2026ctl}.
We compare to a non-parametric analysis presented in \cite{LIGOScientific:2026ctl}, based on binned Gaussian processes (BGP), which makes minimal assumptions imposed on the shape of the distribution \citep{Sridhar2025}, and a  more strongly motivated mixture model analysis based on the methods of \citet{Ray_subpop_2026}.

For BGP, in an earlier analysis, \cite{Sridhar2025} reconstructed the $q$ distribution of the low-mass BBH component defined as systems with primary masses in the range $8$--$25\,M_\odot$, using a data-driven BGP approach applied to GWTC-4.0 \citep{LVK_O4_2025, LIGOScientific:2025slb}. 
Within this mass range, they found a preference for spin orientations aligned with the orbital axis and small spin magnitudes, which can indicate substantial (but not exclusive) contribution from isolated binary evolution. 
These trends have been corroborated in the latest catalog \citep[GWTC-5.0,][]{LIGOScientific:2026ctl}, as well as by alternative statistical models with the latest and earlier catalogs~\citep{Li:2022cul, Godfrey2023,Sadiq:2023zee, Banagiri2025, Godfrey2026,Flanagan:2026ayy}. For our comparison, we use the BGP fits of GWTC-5.0, publicly released by the LVK \citep{LIGOScientific:2026ctl}.

In a more strongly parameterized approach, \cite{Ray_subpop_2026} targeted this preferentially aligned spin and broad $q$ subpopulation in GWTC-4.0 with parametrized mixture models and identified the mass distribution of this subpopulation from the marginal BBH population. 
They found that this population is dominated by low-mass BBH mergers. Several alternate models have found similar results~\citep{Godfrey2023, Li2022, Godfrey2026, Galaudage:2026opk}. 
For our comparison, we reanalyze GWTC-5.0 with the models of \cite{Ray_subpop_2026}.

The comparison between our theoretical results from isolated binary evolution and the two GWTC-5.0 population fits are presented in Figure~\ref{fig:comparison_to_obs}. 
We note that the theoretical models in Figure~\ref{fig:comparison_to_obs} are not statistically fit to the data; they are drawn from our full model grid.
We highlight this subset of models because they most clearly display the features identified in observational analyses: an onset at $q \approxeq 0.4$, with contribution at unequal masses across $q \approxeq 0.6-1.0$, and in some models, a plateau-like feature extending from $q \approxeq 0.7$ to $q = 1$.
As shown in Section~\ref{sec:primary_mass_q_distributions}, the preference for asymmetric $q$ and an equal-mass component are present across the majority of our models;
the selected subset in Figure~\ref{fig:comparison_to_obs} shows a flatter shape between the two features.
One model, M15, results in the equal-mass component dominating and is selected because the non-parametric BGP model, with larger error bars, allows for such a rise to $q\approxeq1$ in addition to a flatter shape between $q\approxeq 0.7-1.0$. 
The selected models here are denoted with a star in the full list of models in Table~\ref{table:models}.
These models span all three SN remnant prescriptions and both BH accretion efficiency prescriptions but cover only $\alpha_\mathrm{CE} = 1$ or $\alpha_\mathrm{CE} = 2$.

Our models are broadly consistent with both distributions inferred from observations. 
The parametric inference of \cite{Ray_subpop_2026} shows a steep rise between $q \approxeq 0.4-0.6$ with a plateau toward $q = 1$, and we find that our models in are excellent agreement across this full range. 
The non-parametric BGP-inferred distribution \citep{Sridhar2025} has large credible intervals that are consistent with our models but do not strongly constrain the shape at higher $q$. 
%Both our models and the inferred distributions show a clear onset near $q \approxeq 0.4$. 
The agreement between our models and the observed low-mass subpopulation emerges across a range of model assumptions.% without fine-tuning to match the observations. 
This demonstrates that the preference for asymmetric $q$ and near-equal-mass contributions are robust structural predictions of isolated binary evolution rather than artifacts of model selection.

\section{Caveats}\label{sec:caveats}
% CE
The treatment of CE evolution in binary population synthesis remains uncertain, and different approaches can affect the relative contribution of CE and SMT to the merging BBH population. 
In \citet{Marchant2021} and \citet{GallegosGarcia2021},
detailed \mesa{} simulations were used with a time-dependent CE calculation in which the star's response to the extreme mass loss and orbital shrinkage at each timestep determined the fate of the CE phase. 
Under this treatment, most systems that survived CE were too wide to merge within a Hubble time, reducing the CE contribution in their grids of simulations. 
The \posydon{} simulations used here apply the two-step $\alpha_\mathrm{CE}$--$\lambda$ prescription (see Section~\ref{sec:methods}), which results in a higher CE contribution than found in \citet{GallegosGarcia2021}. 
The relative contributions of CE and SMT in our results should therefore be interpreted with this in mind. 
However, since both channels contribute to the asymmetric portion of the $q$ distribution, we do not expect our main conclusions about the characteristic features of the $q$ distribution to change significantly.

% HMS-HMS
Our treatment of MT between two hydrogen-rich main-sequence stars leads to very low accretion efficiencies \citep[for examples, see][]{Rocha2024,Zapartas2025}. 
An updated treatment of accretion has been shown to lead to significantly higher MT efficiencies \citep{XingZ2026}. 
This treatment is being implemented into \posydon{} and we expect the $q$ distribution to partially shift closer to unity, bringing our results into even closer agreement with the observed low-mass subpopulation. 
A detailed analysis of the updated treatment on the properties of BBH mergers is forthcoming.

% remnant
Our results are sensitive to both the natal kick and remnant mass prescriptions.
Our SMT results agree with, and extend, \citet{Briel2026}, who used \posydon{} to explore BBH formation through SMT with grids of \mesa{} simulations at eight metallicities. 
While the no-kick models favored a $q$ distribution near one, they found that imposing natal kicks drawn from a Maxwellian distribution with $\sigma = 265\,\mathrm{km\,s^{-1}}$ and scaled inversely with BH mass introduces a low-mass BBH population with unequal $q$.
We show that this low-mass BH population from SMT can contribute significantly to the local population of BBH mergers. 
While the magnitude of natal kicks remains uncertain, there is growing evidence that some stellar-mass BHs must receive at least modest kicks at formation \citep[e.g.,][]{Willems2005,Fragos2009,Kimball2023,Burdge2024,Nagarajan2025}; we therefore use mass-scaled kicks for all models.
The remnant mass prescription sets the minimum BH mass, and thereby informs the location of the asymmetric peak, independently of the kick prescription.
For example, the \citet{Fryer2012} delayed prescription yields a continuous remnant mass function extending to $3-5\ M_{\odot}$; we find that adopting it recovers the same asymmetric and near-equal-mass features, but with the asymmetric peak shifted towards $q\approxeq0.3$ \citep[see also][]{Briel2026}. 
Identifying the kick magnitude at which SMT shifts from equal-mass to asymmetric $q$, and the remnant prescription that sets the exact location of the asymmetric peak, may prove an important diagnostic of the physics of isolated binary evolution.

% secondary peak 
The near-equal-mass contribution in our models arises primarily from systems that evolve through a double-core CE or contact SMT phase, with some contribution from SMT (Gallegos-Garcia et al. 2026, in prep). 
Both of these formation subchannels involve highly uncertain physics: the double-core CE requires modeling two simultaneously overflowing stars within a shared envelope, and contact SMT depends sensitively on the stability criteria for MT between two main-sequence stars. 
Their predicted contributions should therefore be treated with caution.

Finally, we must also consider additional formation mechanisms that may be contributing to the low-mass subpopulation. 
The non-negligible fraction of anti-aligned (negative $\chi_{\mathrm{eff}}$) systems, which recent analyses~\citep{Sridhar2025, Ray:2026qer, Ray_subpop_2026} have constrained to be $\sim20\%$, indicate an abundance of dynamical systems in the low-mass subpopulation~\citep{Tong2025}. Multiple dynamical subchannels such as triple systems, and assembly of first-generation BHs in dense clusters or AGN disks, can be consistent with this anti-aligned fraction. 
Further analysis and additional data are necessary to disentangle the contributions of dynamical formation and isolated binary mergers.
 
\section{Conclusions}\label{sec:conclusions}

We explored the $q$ distribution of merging BBHs formed through isolated binary evolution in the local Universe ($z < 0.5$) using \posydon{}, evolving binary systems from ZAMS to core carbon depletion with detailed \mesa{} simulations. 
We show 15 model configurations varying CE efficiency $\alpha_\mathrm{CE}$, SN remnant mass prescriptions, and BH accretion efficiency.

We find that isolated binary evolution predicts both asymmetric $q$ and a
secondary near-equal-mass contribution, features that persist across a wide
range of model assumptions without fine-tuning; across the models explored here the
asymmetric peak lies near $q \approx 0.5$--$0.7$.
The asymmetric peak reflects the combined contribution of CE and SMT systems, while the near-equal-mass component arises primarily from double-core CE and contact SMT systems. 
These qualitative features are robust despite a wide range in predicted merger rates, demonstrating they are not artifacts of rate-dependent model tuning.

We compared two observational inferences of the low-mass BBH 
subpopulation in GWTC-5.0, a non-parametric BGP reconstruction \citep{Sridhar2025} 
and a parametric mixture model \citep{Ray_subpop_2026}, to our theoretical predictions and find broad consistency with both. 
In particular, our models are in excellent agreement with the parametric inference, which shows a steep rise between $q \approxeq 0.4-0.6$ with a plateau toward $q = 1$. 
Combined with the contrast to the inferred mid- and high-mass subpopulations, which show $q$ distributions strongly inconsistent with our isolated binary evolution models but rather with other formation mechanisms, we conclude that our results provide additional, independent support for the interpretation that the low-mass BBH subpopulation contains a substantial contribution from isolated binary evolution. 

We note that updated treatments of accretion physics between two hydrogen-rich stars increase MT efficiency \citep[e.g.][]{XingZ2026}, and we expect that including this treatment in future studies will shift our predicted asymmetric peak toward more equal $q$, improving agreement with observations further. 
We focused on the general shape of the $q$ distribution and its comparison to observations. 
A detailed investigation of how stellar and binary physics shapes the contributions of each formation subchannel (CE, SMT, double-core CE, and contact SMT) will be presented in a forthcoming study (Gallegos-Garcia et al. 2026, in prep.).

\section*{acknowledgments}
We thank Tassos Fragos and Morgan McLeod for helpful discussions. 
The POSYDON project was supported by the Gordon and Betty Moore Foundation (PI Kalogera, grant award GBMF8477) and the Swiss National Science Foundation (PI Fragos, project No. CRSII5{\_}213497). 
V.K. was partially supported through the D. I. Linzer Distinguished University Professorship fund. 
M.Z. gratefully acknowledges funding from the Brinson Foundation in support of astrophysics research at the Adler Planetarium.
The computations in this paper were partly run on the FASRC Cannon cluster supported by the FAS Division of Science Research Computing Group at Harvard University, 
and partly on the Quest high performance computing facility at Northwestern University, which is jointly supported by the Office of the Provost, the Office for Research, and Northwestern University Information Technology.

\software{\texttt{MESA} \citep{Paxton2011,Paxton2013,Paxton2015,Paxton2019,Jermyn2023};
\citep{townsend2019}
\texttt{Matplotlib} \citep{Hunter2007}; 
\texttt{NumPy} \citep{vanderwalt2011};
\texttt{Pandas} \citep{mckinney-proc-scipy-2010}}

\bibliography{ms}
\bibliographystyle{aasjournal}

\appendix{}

% \section{Relative Contributions from formation subchannels}
% \label{sec:results_subchannel_contribution}

% \begin{figure*}
% \includegraphics[width=0.95
% \textwidth]{SN20_subformationchannels_zhalf_v2}
%     \caption{Contributions of formation subchannels to the $q$ distribution of merging BBHs within the local Universe, $z~<~2.0$. 
%     This model corresponds to \SukhboldNtwenty{} SN prescription with varying \alphaCE{} and MT efficiency.
%     We identify contributions from CE (yellow), double-core CE (dark pink), SMT (light green), contact SMT (dark green), and the full population (gray). 
%     Bottom right panel shows how these models compare to the inferred distributions. }
%     \label{fig:q_channel_contributions_SN20}
    
% \end{figure*}

\begin{table*}
\centering
\caption{Model runs.}
\begin{tabular}{llcc}
\hline\hline
SN Prescription & $\alpha_\mathrm{CE}$ & GRRMHD-informed & Eddington-limited \\
\hline
\multirow{3}{*}{\SukhboldWtwenty{}} 
    & 0.5 & M1  & --   \\
    & 1.0 & M2  & M4   \\
    & 2.0 & M3  & M5$^\star$ \\
\hline
\multirow{3}{*}{\SukhboldNtwenty{}} 
    & 0.5 & M6  & --   \\
    & 1.0 & M7  & M9   \\
    & 2.0 & M8$^\star$  & M10$^\star$ \\
\hline
\multirow{3}{*}{\PattonSukhboldWtwenty{}} 
    & 0.5 & M11 & --   \\
    & 1.0 & M12$^\star$ & M14  \\
    & 2.0 & M13$^\star$ & M15$^\star$ \\
\hline
\end{tabular}
\tablecomments{$^\star$ Models shown in Figure~\ref{fig:comparison_to_obs}.}
\label{table:models}
\end{table*}

\section{Excluding CE via $L_2$ overflow} \label{sec:L2_CE_appendix}

\begin{figure}
    \centering
\includegraphics[width=0.95
\textwidth]{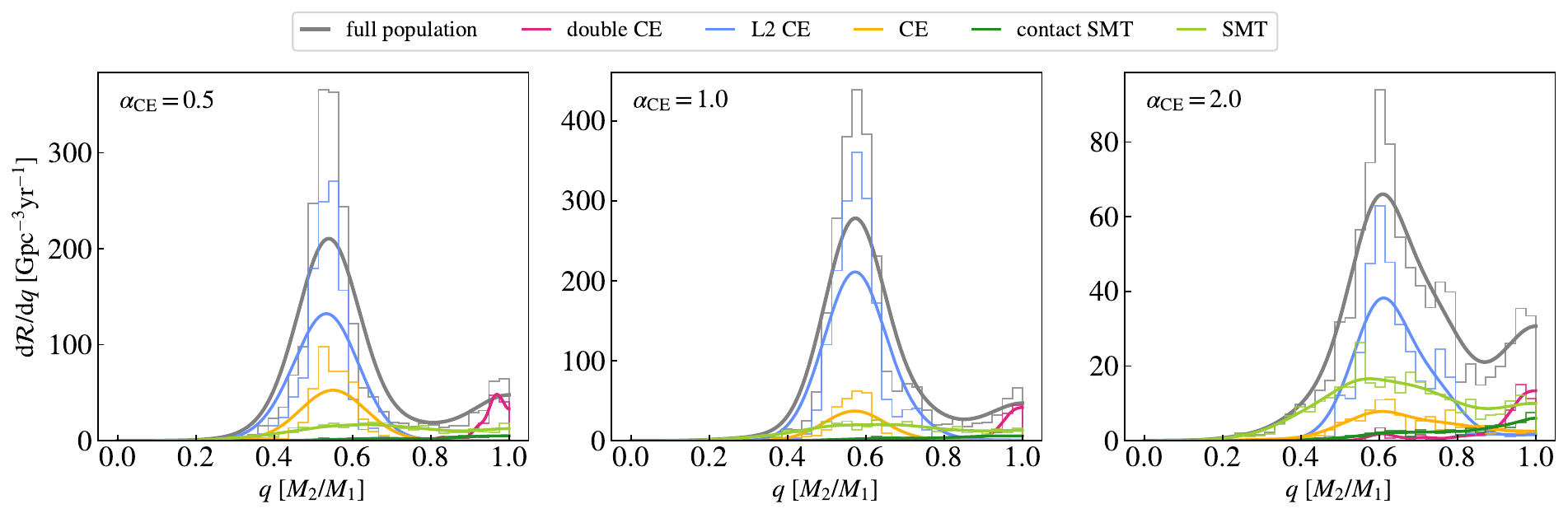}
    \caption{Mass-ratio distributions of distinct formation subchannels of merging BBHs for simulations using the SN remnent prescription of \citep{Sukhbold2016} using engine-N20, including the L2 CE systems. 
    The population of L2 CE has similar features to the CE population. 
    The dominance of the unequal $q$ feature increases when including the L2 CE population.}
    \label{fig:q_with_L2CE_appendix}
\end{figure}

\posydon{} determines the onset of unstable MT triggering a CE phase through several criteria. 
In this section, we describe the subset of CE systems 
triggered by overflow through the outer Lagrangian point, $L_2$, and explain why we exclude them from the main analysis.

The Lagrangian point located behind the accretor is $L_2$ when the accretor is less massive than the donor, and $L_3$ otherwise. 
For the BBH progenitors 
considered here, the accreting BH is typically less massive than its companion at the onset of CE, so outflow occurs through $L_2$.
The conditions under which $L_2$ overflow leads to dynamical instability remain an active area of research \citep[e.g.][]{Sytov2007, Misra2020, MacLeod2018, Lu2023}. 
Material leaving the system through $L_2$ carries higher specific angular momentum relative to standard mass-loss mechanisms, such as isotropic re-emission. 
A significant loss of angular momentum can drive a rapid decrease in orbital separation and decrease the Roche lobe radius around the donor, potentially triggering runaway MT and dynamical instability. 
However, whether $L_2$ overflow inevitably leads to instability is not settled. 
\citet{Marchant2021} used detailed \mesa{} simulations to follow the evolution of binaries experiencing $L_2$ outflow and found that systems can remain stable despite high angular momentum losses, in contrast to treatments that assume instability at the onset of $L_2$ overflow \citep[e.g.,][]{Misra2020}. 
In \posydon{}, we adopt the criterion of \citet{Misra2020}, which treats $L_2$ overflow as unstable when the donor's radius exceeds the volume-equivalent equipotential of $L_2$. 
Given the significant uncertainty in the onset criteria and following evolution, we exclude L2 CE systems from our main results and discuss their properties here.

To identify L2 CE systems, we access the LITE \mesa{} grids underlying the \posydon{} interpolator. 
While the interpolator provides sufficient information for most properties of binary evolution, the termination condition of individual \mesa{} simulations is not interpolated by default so we extract this information from the grid data. 
For each binary system that underwent CE with a BH companion, we identify the nearest-neighbor simulation in the \mesa{} grid using the initial conditions at the onset of the HMS--CO phase, and determine which termination flag was triggered at the end of the simulation. 
Systems flagged as having initiated CE through $L_2$ overflow are separated into a distinct L2 CE category and excluded from the main analysis.

In Figure~\ref{fig:q_with_L2CE_appendix} we show the $q$ distribution of L2 CE systems alongside the other formation subchannels. 
The L2 CE population has similar characteristics to the standard CE population. 
Including L2 CE systems does not change the location of the unequal $q$ peak, but does increase its relative contribution significantly.
In fact, the L2 CE subchannel becomes the dominant subchannel in most of our models when included. 
The range of rates for BBH mergers increases to $\approxeq 70 - 3900\,\mathrm{Gpc}^{-3}\,\mathrm{yr}^{-1}$. 
Given that \citet{Marchant2021} find such systems can remain stable, the default \posydon{} treatment may assign a substantial number of systems to CE that would instead evolve through SMT.

Completely excluding L2 CE systems may be inadvertently removing systems that could have developed to become a BBH merger through other evolutionary pathways, such as the mass loss criteria leading to a CE or through SMT. 
In an effort to explore whether a different $L_2$ treatment would change the population of BBH mergers formed through SMT in \posydon{}, \cite{Briel2026} performed a qualitative comparison to the detailed \mesa{} simulations of \citet{Marchant2021}, which follow BH--H-rich MS binaries through $L_2$ outflow without assuming instability and include the angular momentum losses.
\citet{Briel2026} found that their BH--H-rich systems that undergo L2 overflow do not occupy the orbital period--$q$ parameter space in which \citet{Marchant2021} find BBH mergers forming through SMT (see Figure 6 in \citealt{Briel2026}). 
In other words, further evolution of these systems would not lead to a BBH merger via SMT assuming the results of \citet{Marchant2021}.
The population of BH--H-rich systems does overlap when artificially allowing for fully conservative MT during the HMS--HMS phase. 
The fate of BH--H-rich systems may therefore depend on both the MT phases before and after the formation of the BH.

A thorough treatment of outflow through the outer Lagrangian point requires 3D hydrodynamical simulations.
Nonetheless, given its highly uncertain evolution, it is important to include additional treatments of $L_2$ overflow in \posydon{}, such as that in \cite{Marchant2021} and \cite{Lu2023}. 
This work is underway within the \posydon{} collaboration.

%The population of BH--H-rich systems does overlap when artificially allowing for fully conservative MT during the HMS--HMS phase. 
%The fate of BH--H-rich systems may therefore depend on both the MT phases before and after the formation of the BH.
%We conclude that although $L_2$ overflow remains an uncertain mechanism in binary evolution, we do not expect our exclusion of L2 CE systems to significantly affect the key features identified for the subchannels in isolated binary evolution.

\end{document}